\documentclass[twoside,showpacs,superscriptaddress,twocolumn,floatfix,a4paper,aps,prl,longbibliography]{revtex4-1}

\usepackage{url}
\usepackage[colorlinks]{hyperref}
\hypersetup{%
    plainpages=true,
    breaklinks=true,
    hypertexnames=false,
    pageanchor=true,
    colorlinks=true,
    linkcolor={blue},
    citecolor={red},
    urlcolor={blue},
    anchorcolor={black}
}

\usepackage{color}
\usepackage{graphicx}
\usepackage[utf8x]{inputenc}
\usepackage[T1]{fontenc}
\usepackage{amstext}
\usepackage{amsmath}
\usepackage{hyperref}
\usepackage{dcolumn}

\newcommand\ket[1]{|#1\rangle}
\newcommand\bra[1]{\langle #1|}
\newcommand\oprod[2]{\ket{#1}\bra{#2}}

\newcommand\SI[2]{#1~#2}

\begin{document}

\title{Bell nonlocality and fully-entangled fraction measured
in an entanglement-swapping device without quantum state
tomography}

\author{Karol Bartkiewicz} \email{bark@amu.edu.pl}
\affiliation{Faculty of Physics, Adam Mickiewicz University,
PL-61-614 Pozna\'n, Poland} \affiliation{RCPTM, Joint Laboratory
of Optics of Palack\'y University and Institute of Physics of
Czech Academy of Sciences, 17. listopadu 12, 771 46 Olomouc, Czech
Republic}

\author{Karel Lemr}
\email{k.lemr@upol.cz} \affiliation{RCPTM, Joint Laboratory of
Optics of Palack\'y University and Institute of Physics of Czech
Academy of Sciences, 17. listopadu 12, 771 46 Olomouc, Czech
Republic}

\author{Antonín Černoch} \email{acernoch@fzu.cz}
\affiliation{Institute of Physics of the Czech Academy of
Sciences, Joint Laboratory of Optics of PU and IP AS CR, 17.
listopadu 50A, 772 07 Olomouc, Czech Republic}

\author{Adam Miranowicz} \email{miran@amu.edu.pl}
\affiliation{Faculty of Physics, Adam Mickiewicz University,
PL-61-614 Pozna\'n, Poland}

\date{\today}

\begin{abstract}
We demonstrate an efficient experimental procedure based on
entanglement swapping to determine the Bell nonlocality measure of
Horodecki et al. [Phys. Lett. A {\bf 200}, 340 (1995)] and the
fully-entangled fraction of Bennett et al. [Phys. Rev. A {\bf 54}, 3824
(1996)] of an arbitrary two-qubit polarization-encoded state. The
nonlocality measure corresponds to the amount of the violation of
the Clauser-Horne-Shimony-Holt (CHSH) optimized over all
measurement settings.
By using simultaneously two copies of
a given state, we measure directly only six parameters. Our method
requires neither full quantum state tomography of 15 parameters
nor continuous scanning of the measurement bases used by two
parties in the usual CHSH inequality tests 
with four measurements in each optimization step.  
We analyze how well the measured degrees
of Bell nonlocality and other entanglement witnesses (including
the fully-entangled fraction and a nonlinear entropic witness) of
an arbitrary two-qubit state can estimate its entanglement. In
particular, we measured these witnesses and estimated the
negativity of various two-qubit Werner states. Our approach could
especially be useful for quantum communication protocols based on
entanglement swapping.
\end{abstract}

\maketitle

Experimental methods for detecting and quantifying quantum
entanglement~\cite{Schroedinger35,EPR35} and Bell nonlocality
(usually identified with the violation of a Bell
inequality)~\cite{Bell64,Bell} are of paramount importance for
practical quantum-information processing~\cite{NielsenBook},
quantum cryptography (e.g., quantum key
distribution)~\cite{Ekert91PRL}, and quantum communication (e.g.,
quantum teleportation)~\cite{Bennett93PRL}. The importance of this
topic can be highlighted by the fact that the world's first
quantum satellite, emitting pairs of entangled photons, has
recently been launched~\cite{quantSat}. Since the seminal
experiments of Aspect \emph{et al.}~\cite{Aspect81PRL,
Aspect82aPRL,Aspect82bPRL} in the early 1980s, various methods of
detecting entanglement and nonlocality have been developed (for
reviews see~\cite{Horodecki09RMP, Brunner14RMP}). Note that only
very recently loophole-free tests of Bell nonlocality have been
performed~\cite{Hensen15Nature, Shalm15PRL}. Nevertheless, to
measure a degree of these effects seems to be much more difficult
and important rather than only to detect them.

Thus, the question arises how to determine some entanglement or
nonlocality measures, e.g., for only two qubits. These can
include: (i) the negativity $N$, related to the Peres-Horodecki
inseparability criterion~\cite{Peres96PRL,Horodecki96PLA}, which
is a measure of the entanglement cost under the operations
preserving the positivity of the partial transpose of a
state~\cite{Audenaert03PRL,Ishizaka04PRA}; moreover, the
negativity is an estimator of entanglement dimensionality, i.e.,
the number of the entangled degrees of freedom of two
subsystems~\cite{Eltschka13PRL};  (ii) the concurrence $C$,
corresponding to the entanglement of
formation~\cite{Wootters98PRL}; or (iii) the Bell nonlocality
measure $B$ of
Ref.~\cite{Horodecki95PLA,Horodecki96aPLA,Horst13PRA,Bartkiewicz13PRA_CHSH}
corresponding the violation of the Bell inequality derived by
Clauser, Horne, Shimony, and Holt (CHSH)~\cite{CHSH69PRL}, which
is optimized over all measurements (i.e., detector settings) on
sides A (Alice) and B (Bob). We note that these measures are
equivalent as $N=C=B$ for, e.g., entangled pure states and
these states subjected to phase damping (i.e., a special kind of
Bell diagonal states)~\cite{Miranowicz04PLA, Horst13PRA}.

One could argue that the most straightforward experimental method
for quantifying entanglement and nonlocality is to perform a
complete quantum-state tomography (QST) to determine a given
bipartite state $\rho$ and, then, to calculate (from $\rho$) its
entanglement and nonlocality measures related to a specific
quantum-information task. However, for the simplest case of two
qubits in a general mixed state $\rho$, at least 15 (types of)
measurements should be performed on identical copies of $\rho$ to
determine all 16 real parameters of $\rho$. Now the question
arises whether a measure of entanglement or nonlocality could be
determined directly or at least by a smaller number of
measurements corresponding to an incomplete QST.

Various theoretical proposals to efficiently detect and quantify
entanglement and nonlocality were described in, e.g.,
Refs.~\cite{Horodecki02PRL,Horodecki03PRL,Mintert04PRL,
Mintert05PRL,Carteret05PRL, Aolita06PRL, Bartkiewicz13PRA_CHSH,
Bartkiewicz15PRA_W}. The first experimental direct measurement of
a nonlinear entanglement witnesses was reported
in~\cite{Bovino05PRL}. While the first experimental determination
of an entanglement measure (i.e., the concurrence, being equal to
the negativity and the CHSH measure) for a two-qubit pure state
was reported in Ref.~\cite{Walborn06Nat}. Probably, the first
experimental method for measuring a collective universal witness,
as a conclusive entanglement detection of any two-qubit mixed
state (encoded in photon polarization), was proposed in
Ref.~\cite{Bartkiewicz15PRA_W}. Unfortunately, this method is much
more complicated than QST and requires to erase some measured
information to ensure its optimality. All these theoretical and
experimental studies show the fundamental difficulties not only in
quantifying, but even in conclusively detecting the entanglement
and nonlocality of a two-qubit state without QST.

The CHSH inequality has been mostly used for detecting and
quantifying~\cite{Bartkiewicz13PRA_CHSH} Bell nonlocality of two
qubits. This can be done by determining, e.g., the nonlocality
measure $B$ corresponding to finding an optimal set of
measurements for the correlated subsystems. If one deals with an
unknown state, this approach requires applying all possible
two-measurement settings for each qubit to find the optimal ones.
However, as shown in~\cite{Bartkiewicz13PRA_CHSH}, a more direct
experimental procedure, which requires using only six-detector
settings, can be applied to find the maximal violation of the CHSH
inequality for an arbitrary unknown two-qubit state. To avoid
implementing inefficient procedures to be optimized over all
possible measurement bases, this alternative approach of
Ref.~\cite{Bartkiewicz13PRA_CHSH} requires using simultaneously
two copies of a given two-qubit state.   The estimation of the
amount of entanglement from the maximum violation of the CHSH
inequality was studied in, e.g.,~\cite{Bartkiewicz13PRA_CHSH}.

In this paper, we experimentally implement a direct and efficient
method to conclusively detect Bell nonlocality and to determine
the CHSH measure for two-qubit mixed state without QST. In
particular, for phase-damped two-qubit pure states, our method
reduces to determining the concurrence and negativity.

Specifically, we report here the experimental implementation of
our six-step measurement procedure for determining the Bell
nonlocality measure $M$ [where $B=\sqrt{\max(M,0)}$], defined in
Eq.~(\ref{eq:M}), with two copies of the investigated state and
 the singlet-state projection implemented by Hong-Ou-Mandel
(anti-)coalescence (see, e.g.,
Refs.~\cite{HOM,Bovino05PRL,Bartkiewicz13corr,Bartkiewicz13PRA_CHSH,Bartkiewicz13sfid,tomo1,tomo2}).
In our experiment, we use polarization-encoded qubits. Our
approach utilizes only a single two-photon interference event,
instead of two required for standard nonlinear
approaches~\cite{Bovino05PRL,Xu14PRL}. However, we are able to
measure the same nonlinear entropic entanglement witness, as in
Refs.~\cite{Bovino05PRL,Xu14PRL}, for subsystems of equal purity.
Here we also measure a more sensitive entanglement witness, i.e.,
the fully-entangled fraction (FEF) $f$ of a two-qubit state
$\rho$, which is defined as~\cite{Bennett96}:
$f(\rho)=\max_{\ket{e}}\bra{e}\rho\ket{e}$, where the maximum is
taken over all maximally-entangled states $\ket{e}$. The FEF
has been shown to be a useful concept in describing realistic QIP
protocols including dense coding, teleportation, entanglement
swapping, quantum cryptography based on Bell's inequality and, in
general, multiqubit entanglement (see,
e.g.,~\cite{Bennett96,Horodecki99,Albeverio02,Zhou02,
Grondalski02PLA, Ozdemir07PRA,Li08PRA,Lee09PRA, Zhao10JPA,Li12QIC,
Zhao15PRA}). The experimental complexity of our method of
measuring the optimal CHSH inequality violation and the FEF is
comparable to that of measuring the collectibility witness of
Refs.~\cite{Rudnicki11PRL, Rudnicki12PRA, Lemr16collect} and can
be implemented with the same experimental resources. Note that,
contrary to the FEF and nonlocality measure $M$, the usefulness of
the collectibility witness is limited mainly to pure or almost
pure states only~\cite{Rudnicki11PRL, Rudnicki12PRA,
Lemr16collect}. In addition to measuring the Bell nonlocality
measure, we can apply the same method to measure the maximum
achievable fidelity and FEF.

\begin{figure*}
\includegraphics[width=5.7cm]{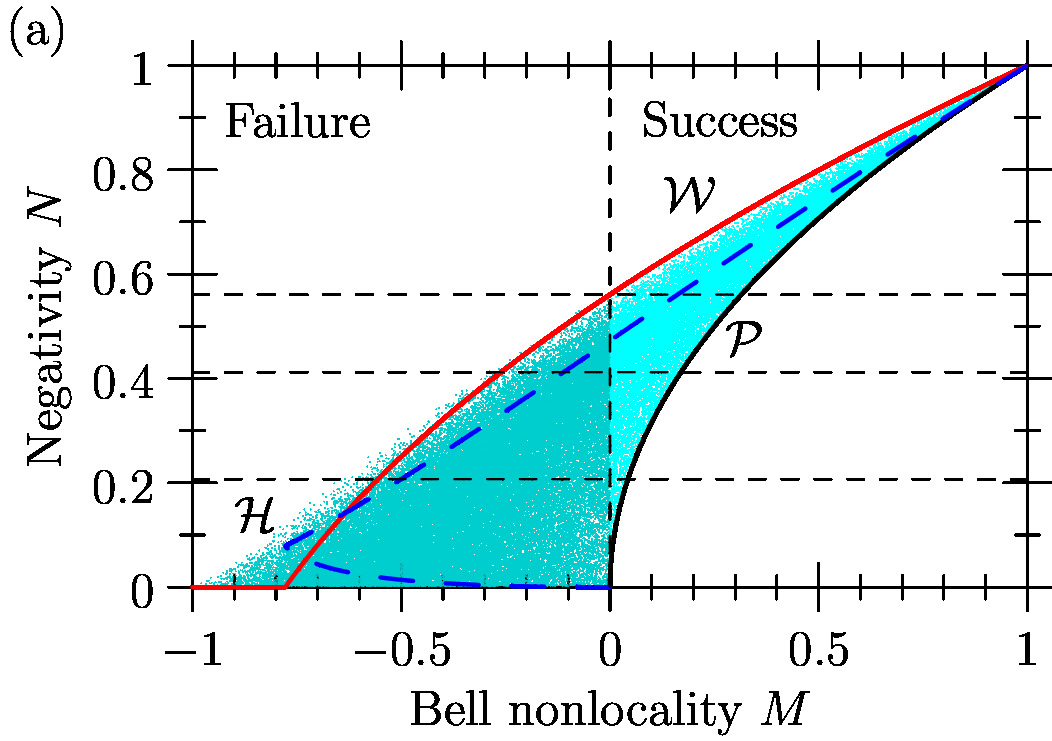}
\includegraphics[width=5.7cm]{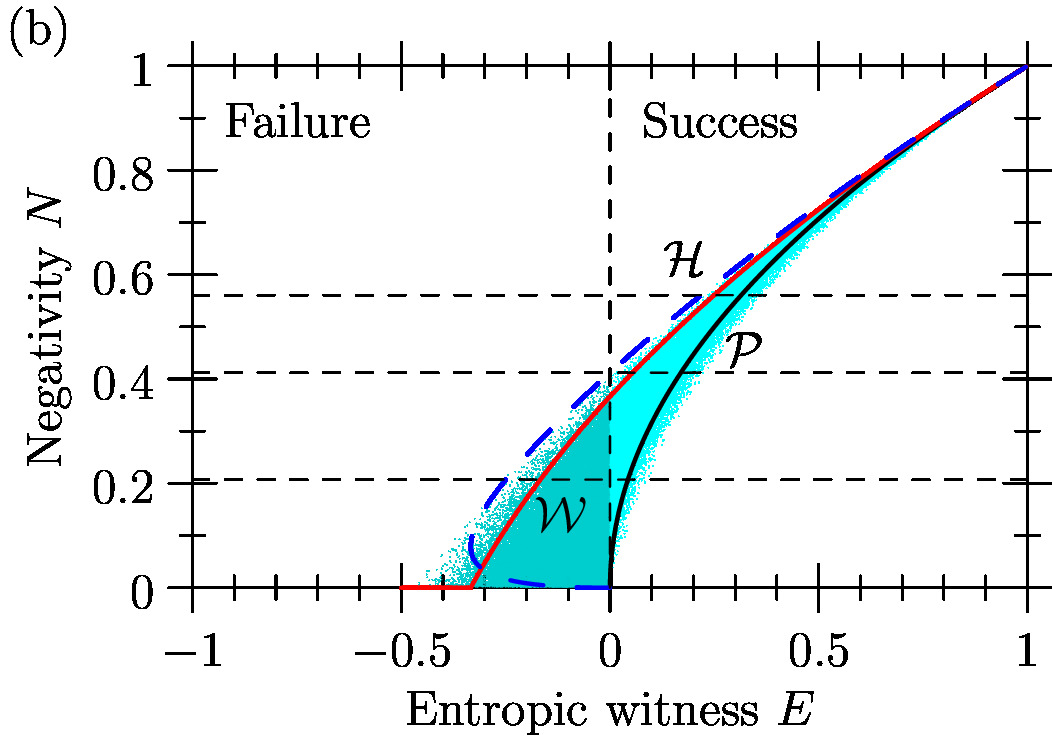}
\includegraphics[width=5.7cm]{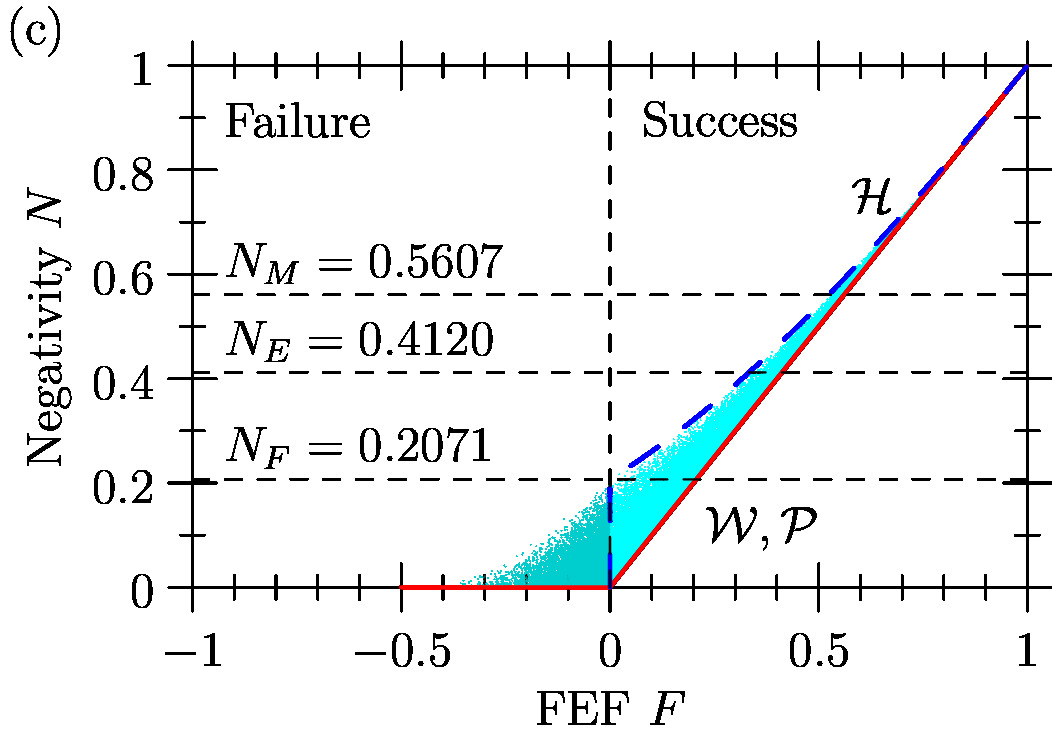}
\caption{\label{fig:num} (Color online) Amount of entanglement
measured with the negativity 
$N$~\cite{Negativity1,Negativity2,Horodecki09RMP} versus (a) the
Bell nonlocality measure $M$,  defined in  Eq.~(\ref{eq:M}),
(b) the entropic witness $E$,  given in Eq.~(\ref{eq:E}), and
(c) the FEF $F$, defined in Eq.~(\ref{eq:X}), for $10^5$ two-qubit
states randomly generated by a Monte Carlo simulation.
Entangled states for which an entanglement witness is successful
in detecting inseparability are marked with light cyan dots. The
entangled states that are ignored by the respective witness are
marked with dark cyan dots. The Werner ($\mathcal{W}$),
Horodecki ($\mathcal{H}$), and pure ($\mathcal{P}$) states,
defined in Eq.~(\ref{eq:states}), correspond to the upper solid,
dashed, and lower solid curves, respectively. In particular, $F$
allows detecting the entanglement of the Werner states
$\mathcal{W}$ for the whole range of their mixing parameter
$p>1/3$. A given witness detects all the entangled states of the
negativity above its respective threshold, i.e., $N_M=0.5607,$
$N_E=0.4120,$ and $N_F=0.2071.$ }
\end{figure*}

\vspace*{3mm} \noindent {\it Theoretical framework.---} In our
experiment we study photonic qubits encoded in polarization of
single photons. The associated Pauli matrices are
defined as $\sigma_1 = \oprod{D}{D}-\oprod{A}{A},$ $\sigma_2 =
\oprod{L}{L}-\oprod{R}{R},$ $\sigma_3 =
\oprod{H}{H}-\oprod{V}{V},$ where each capital letter corresponds
to a particular polarization direction state (i.e., $D$ for
diagonal, $A$ for antidiagonal, $L$ for left-circular, $R$ for
right-circular, $H$ for horizontal, and $V$ for vertical
polarizations). A general two-qubit state can be expressed in the
Hilbert-Schmidt form as
\begin{equation} \rho = \frac{1}{4}(I\otimes
I+\vec{x}\cdot\vec{\sigma}\otimes
I+I\otimes\vec{y}\cdot\vec{\sigma}+\!\!\!\sum
\limits_{i,j=1}^{3}T_{i,j}\,\sigma _{i}\otimes \sigma _{j}),
\label{eq:rho}
\end{equation}
where $\vec{\sigma}=[\sigma_{1},\sigma_{2},\sigma_{3}]$.
The elements of the Bloch vectors read as
$x_{i}=\mathrm{Tr[}\rho(\sigma_{i}\otimes I)]$ and
$y_{i}=\mathrm{Tr[}\rho(I\otimes\sigma_{i})]$), respectively.
Finally, the correlation matrix $T$ is defined as
$T_{i,j}=\mathrm{Tr}[\rho(\sigma_{i}\otimes\sigma_{j})]$
for $i,j=1,2,3.$

{The CHSH inequality for a two-qubit state
$\rho\equiv\rho_{ab}$ can be written
as~\cite{CHSH69PRL,Horodecki09RMP}:
$|\mathrm{Tr}\,(\rho \,{\mathcal B}_{\mathrm{CHSH}})|\leq 2.$
The maximum possible average value of the CHSH
operator~\cite{Horodecki95PLA} is
\begin{equation}\label{eq:CHSH}
\max_{{\mathcal B}_{\mathrm{CHSH}}}\,\mathrm{|Tr}\,(\rho
\,{\mathcal B}_{ \mathrm{CHSH}})|=2\,\sqrt{g(\rho )},
\end{equation}
where $\mathcal{B}_{\mathrm{CHSH}}={ \hat{a}\cdot\vec\sigma
\otimes(\hat{b}+\hat{b}')\cdot \vec\sigma +
a'\cdot\vec\sigma\otimes(\hat{b}-\hat{b}')\cdot \vec\sigma}$
depends on  real unit vectors $\hat a, \hat b,\hat a', \hat
b'$. The function $g(\rho)=\mathrm{Tr}R- \min[\mathrm{eig}(R)]\leq
2$} depends on the eigenvalues $\mathrm{eig}(R)$ of a real
symmetric matrix $R\equiv T^{T}\,T,$ which is described with only
six parameters, e.g., $R_{i,j}$ with $i\ge j.$ As shown
in~\cite{Bartkiewicz13PRA_CHSH}, these six elements can be
measured directly using two copies of $\rho$ (i.e., $\rho_1$ and
$\rho_2$). This is a consequence of the following identity
\begin{equation}
R_{i,j}=\mathrm{Tr}\left[(\rho_{a_1b_1}\otimes
  \rho_{a_2b_2})S_{a_1a_2}\otimes(\sigma_i\otimes\sigma_j)_{b_1b_2}\right] ,\label{eq:R}
\end{equation}
where $\rho_{a_1b_1}\equiv\rho_1$ and $\rho_{a_2b_2}\equiv\rho_2$
for the subsystems $a$ and $b$, whereas the operator $S_{a_1a_2} =
(I-4| \Psi^-\rangle\langle\Psi^-| )_{a_1a_2}$ is given in terms of
the singlet state $|\Psi^-\rangle = (\ket{HV}-\ket{VH})/\sqrt{2}$
and the two-qubit identity operation $I$. As follows from
Eq.~(\ref{eq:CHSH}), the CHSH inequality $|\mathrm{Tr}\,(\rho
\,{\mathcal B}_{\mathrm{CHSH}})|\leq 2$ is violated  iff
$f(\rho)>1$. Here we apply the Horodecki measure of Bell (or CHSH)
nonlocality defined as~\cite{Horodecki96PLA}: 
\begin{equation}\label{eq:M}
M=g-1=\mathrm{Tr}R - \min[\mathrm{eig}(R)]-1,
\end{equation}
which is positive  iff the CHSH inequality is violated and
reaches its maximum $M=1$ for maximally-entangled states.
Note that $M$ is trivially related to the measure of Bell
nonlocality $B=\sqrt{\max(M,0)}$ studied
in~\cite{Miranowicz04PLA,Horst13PRA,Bartkiewicz13PRA_CHSH}.
Moreover, we apply another entanglement witness, i.e. the
(modified) FEF $F(\rho)$ defined as
\begin{equation}\label{eq:X}
F=2f-1=\tfrac{1}{2}\left(\mathrm{Tr}\sqrt{R} -1\right),
\end{equation}
which is a rescaled version of the standard FEF
$f(\rho)$~\cite{Horodecki96PRA,Badziag00PRA,Grondalski02PLA}. Note
that $F<0$ for all separable states and equals to the negativity
for the Werner and pure states (see Fig.~\ref{fig:num}c). These
FEFs correspond to the fidelity of various entanglement-assisted
processes maximized over all possible local unitary operations.
The FEF $F$ detects more entangled states than both nonlocality
measure $M$ and another nonlinear entropic witness, measured in
Ref.~\cite{Bovino05PRL} and defined by
\begin{eqnarray}
E&=&\mathrm{2(Tr}\rho_{ab}^2\nonumber
- \min[\mathrm{Tr}\rho^2_a,\mathrm{Tr}\rho^2_b])\\
&=&\tfrac{1}{2}(\mathrm{Tr}R+|\mathrm{Tr}\rho^2_a-\mathrm{Tr}\rho^2_b|-1),\label{eq:E}
\end{eqnarray}
if considered separately (see Fig.~\ref{fig:num}). Note that
$E=\tfrac{1}{2}(\mathrm{Tr}R-1)$ for
$\mathrm{Tr}\rho^2_a=\mathrm{Tr}\rho^2_b$. The spectrum of $R$,
used in the definition of $F$, is measured unavoidably while
measuring $M=g(\rho)-1$, which quantifies the optimal CHSH
violation. Thus, the optimal CHSH inequality is fundamentally more
powerful in detecting quantum entanglement than its original
form in an unoptimized measurement basis.

The performance of a given entanglement witness can conveniently
be studied with one-parameter ($0\le p\le 1$) classes of states
(see Fig.~\ref{fig:num}) including the Werner states
$\mathcal{W}$~\cite{Werner89PRA,Miranowicz04PLA}, the Horodecki
states $\mathcal{H}$~\cite{Horodecki96PLA}, and pure states
$\mathcal{P}$ defined, respectively, as
\begin{eqnarray}\nonumber
\mathcal{W}(p)&=&\tfrac{(1-p)}{4}I + p\ket{\Psi^-}\bra{\Psi^-},\\\nonumber
\mathcal{H}(p)&=&p\ket{HH}\bra{HH} + (1-p)\ket{\Psi^-}\bra{\Psi^-},\\
\mathcal{P}(p)&=&(\sqrt{p}\ket{HH}+\sqrt{1-p}\ket{VV})(\mathrm{H.c.}),\label{eq:states}
\end{eqnarray}
where H.c. stands for the Hermitian conjugate of the
preceding terms.

\begin{figure}
\includegraphics[width=9cm]{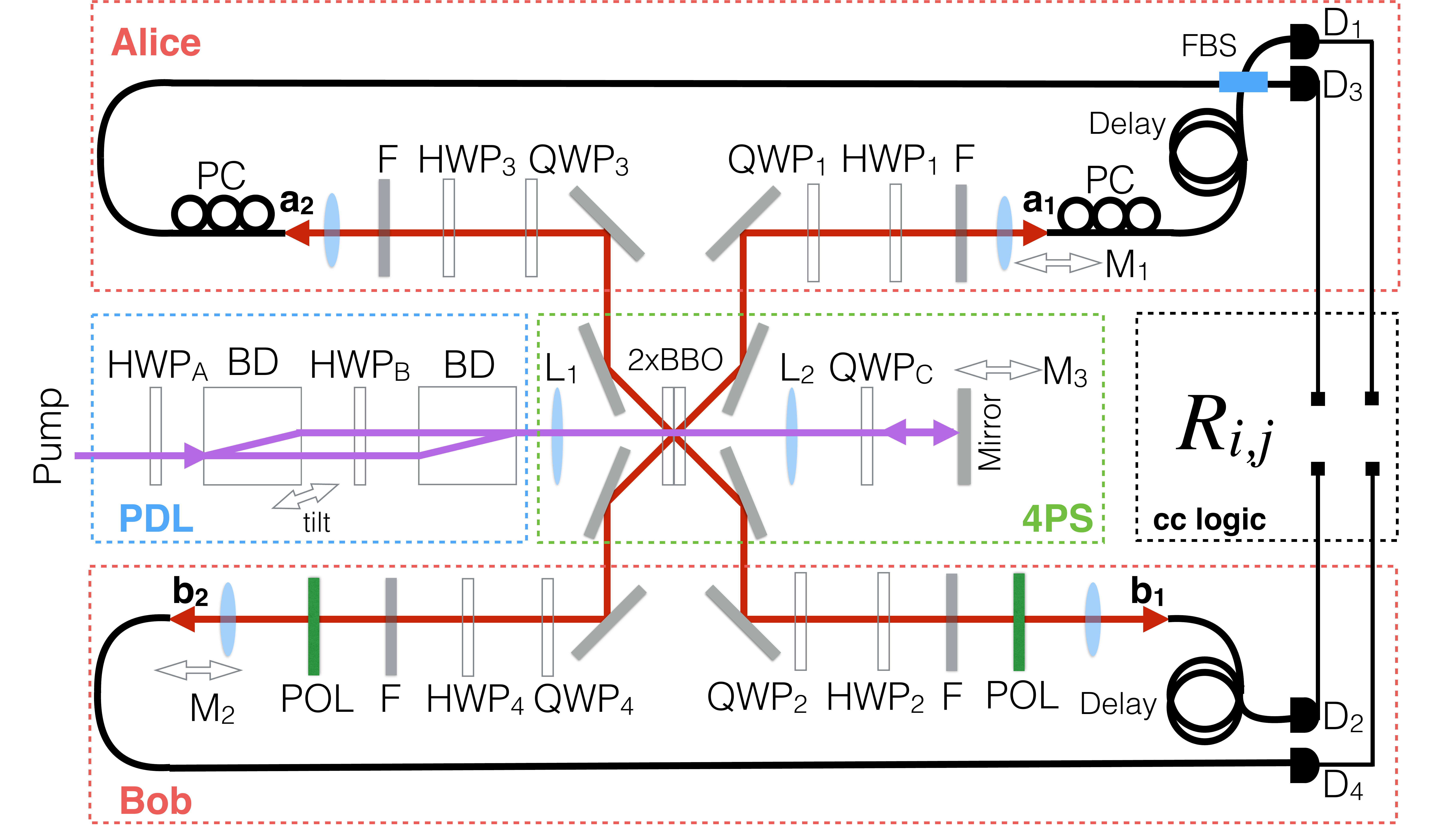}
\caption{\label{fig:setup} (Color online) Experimental scheme for
determining the Bell nonlocality measure $M$ and the FEF $F$
of polarization-encoded two-qubit states with linear optics via
the elements of the $R$ matrix, defined in Eq.~(\ref{eq:R}). This
setup consists of narrow-band filters (Fs), half-wave plates
(HWPs), quarter-wave plates (QWPs), beam dividers (BDs), detectors
(Ds), a fibre beam splitter (FBS), polarization controllers (PCs),
lenses (Ls), BBO crystals, mirrors, and  motorized-translation
stages (M). Note that this is an entanglement-swapping
device, where the swapping is implemented by the FBS. The setup
is powered by a laser system described in the text. The
polarization-dispersion line (PDL) compensates the polarization
dispersion introduced by the BBO crystals in the
four-photon-source module (4PS). In this module, two copies of a
two-qubit state ${\rho}_1$ and ${\rho}_2$ are prepared in the
modes $(a_1,\,b_1)$ and $(a_2,\,b_2)$, respectively. We name the
last two modules as belonging to Alice and Bob, respectively. In
Alice's module, qubit $a_1$ is overlapped on an 50:50 beam
splitter with qubit $a_2$ to implement the measurement of
$S_{a_1a_2} = (I-4|\Psi^-\rangle\langle\Psi^-| )_{a_1a_2}$. In
Bob's module, qubits $b_1$ and $b_2$ are projected onto the
eigenstates of $(\sigma_i\otimes\sigma_j)_{b_1b_2}$ for $i\ge j$
and $i,j=1,2,3$ by the respective polarizer (POL) and detected at
the respective detector. The four-fold coincidence counts are then
processed to estimate the values of $R_{i,j}$. }
\end{figure}

\vspace*{3mm} \noindent {\it Experiment.---} In our experiment we
used a four-photon source shown in Fig.~\ref{fig:setup}. This
multiphoton source is pumped by the Coherent Mira femtosecond
laser at repetition rate of \SI{80}{MHz}. The wavelength of the
pulses is then converted in the process of second-harmonic
generation (SHG) to \SI{413}{nm}. On average the mean power of the
up-converted pumping beam is circa \SI{300}{mW}. Next, the beam
travels through a polarization-dispersion line (PDL) that
compensates the polarization dispersion caused by the
$\beta$-BaB$_2$O$_4$ crystals (BBO) used to create pairs of
photons. The PDL was build by placing a half-wave plate (HWP)
between two beam displacers (BDs). This construction allows us to
tune the relative optical path of photons of selected polarization
by tilting the BDs. The pumping beam then powers a BBO crystal
cascade~\cite{Kwiat99PRA}, which generates (in the process of
type-I spontaneous parametric down-conversion) pairs of
horizontally- and vertically-polarized photons. The polarization
and phase of a single photon pair can be adjusted by setting the
correct polarization of the pumping beam. The beam passes through
a quarter-wave plate (QWP) before and after being reflected by a
mirror. This QWP compensates the polarization dispersion in the
BBO crystals, which are now pumped in the opposite direction and
create a second pair of photons.

The created pairs of photons are reflected by axillary mirrors to
Alice and Bob who process the relevant photons $(a_1,a_2)$
and $(b_1,b_2)$ from each pair, respectively. The polarizations
of photons $b_1$ and $b_2$ are first rotated by QWPs and HWPs and
then projected by polarizers (POLs) to match an eigenstate of
$(\sigma_m\otimes\sigma_n)_{b_1b_2}$. Next, the photons are
coupled to single-mode fibers and detected. Photons $a_1$ and
$a_2$ are coupled to fibers directly, and then overlapped on a
balanced fiber beam splitter (FBS) before being detected. Before
entering the fibers photons $a_1$ and $a_2$ ($b_1$ and $b_2$) are
filtered with \SI{5}{nm} (\SI{10}{nm}) interference filters.
Note that entanglement swapping in our setup can be
implemented by the FBS.

\begin{figure}
\includegraphics[width=8.5cm]{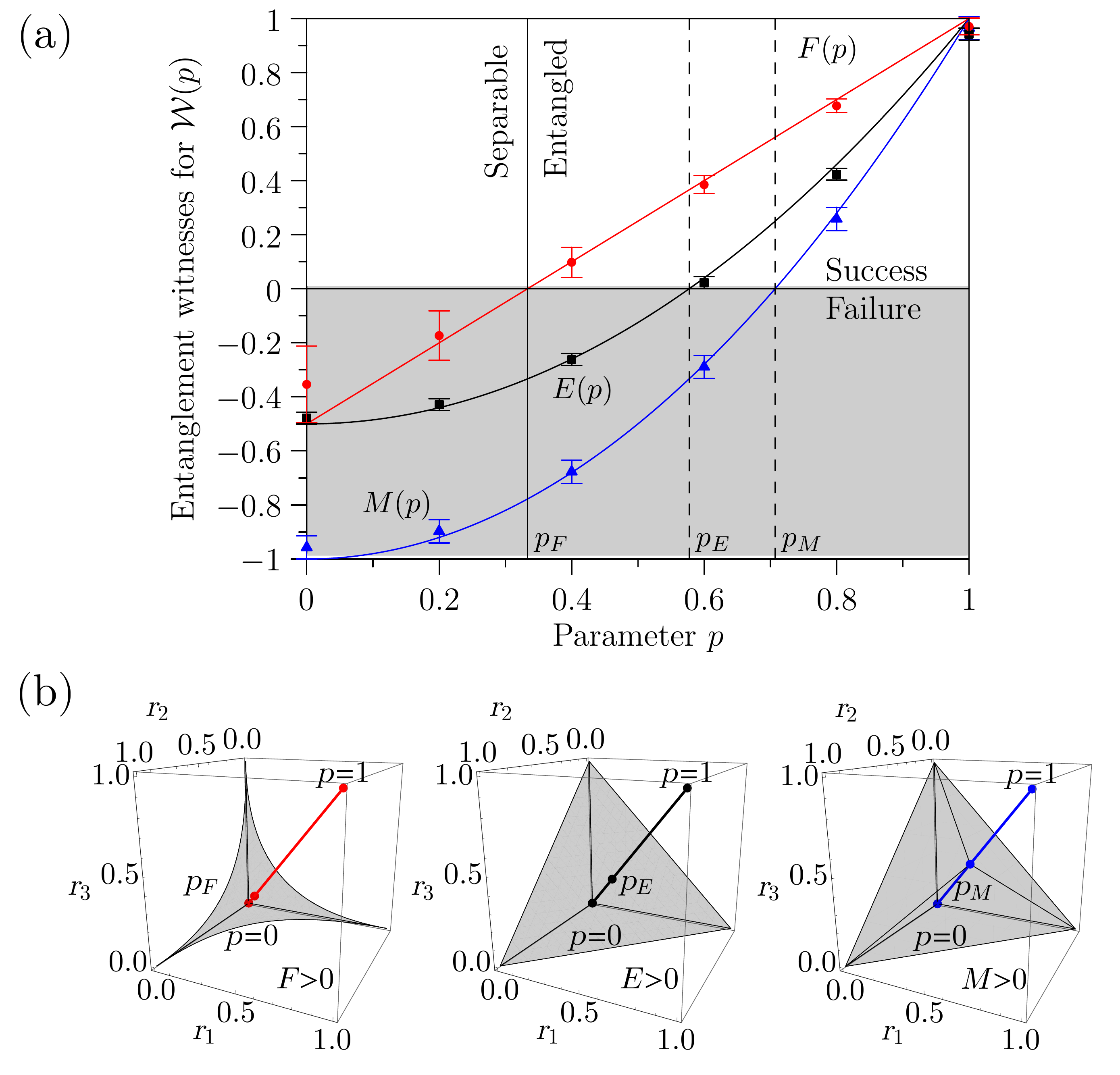}
\caption{\label{fig:werner} (Color online) (a) The Bell
nonlocality measure $M$, the FEF $F$ and the entropic witness $E$
versus the mixing parameter $p$ of the Werner states
$\mathcal{W}$.  Our theoretical results for the perfect Werner
states are marked with solid curves. The systematic deviation from
the ideal case (solid curves) is caused by the fact that our 
experimentally-created singlet state was not ideally pure and its
purity reached circa $93\%$. Moreover, our 
experimentally-created mixed state was not totally mixed. The
ideal Werner are separable for $p<1/3$. For these states the
entanglement can be detected with the FEF $F$, entropic witness
$E$, and Bell nonlocality $M$ for $p_F>1/3,\,p_E>1/\sqrt{3},$ and
$p_M>1/\sqrt{2},$ respectively. The separable states enclosed by
gray boundaries for the respective witnesses are shown in panel
(b), where $[r_1,r_2,r_3]=\mathrm{eig}(R)$ and the Werner states
are located on the diagonals. The entanglement is detected, if the
respective witness is nonnegative. }
\end{figure}

The interference strength on the FBS is tuned by a proper choice
of a fiber delay and by setting the right position of the 
motorized-translation stage (M) associated with the
corresponding mirror in a four-photon source (4PS). For its two
extreme settings, the FBS performs the projective
measurements $I/2$ or $|\Psi^-\rangle\langle\Psi^-|$. However, the
optical couplers collect photon pairs generated at random
distances from each other in the BBO crystal due to its 
group-velocity dispersion. Thus, a fraction of photons $r$ will
not overlap on the FBS, but can be detected in the same time
window of the detectors as the perfectly-overlapping photons,
i.e., Alice performs $[\frac{r}{2}I +(1-r)
|\Psi^-\rangle\langle\Psi^-|]_{a_1a_2}$ measurement. For each
source configuration, we measure this fraction of noninteracting
photons while calibrating the setup and setting the appropriate
delays. Depending on the weight $r$, Alice performs a
projection on a particular Werner state. Thus, the uncertainty of
the obtained results is limited only by the number of the
registered coincidences and the precision of determining the
weight $r$. In this setup, we typically register one
four-fold coincidence event in 5 minutes. We collected hundreds of
such coincidences per a measurement setting. In our experiment we
experimentally studied two kinds of two-qubit states, namely the
pure separable states $\mathcal{P}(0)=\ket{VV}\bra{VV}$, and the
Werner states, defined in Eq.~(\ref{eq:states}), which can
be entangled even if $M<0$ (see Fig.~\ref{fig:werner}). In
particular, we measured the maximally-entangled states
$\mathcal{W}(1)$ and the completely-mixed state $\mathcal{W}(0)$.
These states were prepared using the method described in
Ref.~\cite{Lemr16collect}. The Werner states are particularly
important for quantum technologies because entanglement
purification schemes transform other states into the Werner
states~(see Ref.~\cite{Pan01Nat} and the references therein). Our
measurement results for the Werner states are summarized in
Fig.~\ref{fig:werner}. In all these cases, we reconstructed
matrices $R$ and applied the maximum likelihood method to estimate
their spectra. For the remaining experimental results and
technical details see the Supplement~\cite{Supplement}.

\vspace*{3mm}
\noindent {\it Conclusions.---} By applying the maximum likelihood
method, we demonstrated that direct measurements of nonlinear
entanglement witnesses~\cite{Bovino05PRL,Bartkiewicz15PRA_W} could
be made robust to experimental errors by exploiting the
correlations between them. Our procedure is applicable if the mean
experimental matrix $R^{(\mathrm{exp})}$ can be well approximated
with its maximum likelihood estimate $R$ ~\cite{Supplement}. This method
could further be applied to improve the error robustness of
entanglement measures~\cite{Bartkiewicz15PRA_N}.

Our Monte Carlo simulations also allowed us to compare the
efficiency of our entanglement detection by means of the Bell
nonlocality measure $M$, the FEF $F$, and the entropic witness
$E$ based on double Hong-Ou-Mandel interference, as described
Ref.~\cite{Bovino05PRL}. We measured the FEF $F$ that is an
entanglement witness more powerful in detecting entanglement
than $M$ and $E$. This was exemplified by our study of the Werner
states. These states are recognized to be entangled by a
particular entanglement witness if they have large enough value of
the mixing parameter $p$ (see Fig.~\ref{fig:werner}). For
the FEF $F$, this critical value is $p_F>\tfrac{1}{3}$,
which corresponds to the range for which the Werner states are
entangled. For the entropic witness $E$ and Bell nonlocality
measure $M,$ the entanglement of the Werner states occurs if
$p_E>1/\sqrt{3}$ and $p_M>1/\sqrt{2},$ respectively.

In our experiment, we have conducted a conclusive Bell nonlocality
test by means of two-photon pairs prepared in the same state and
six independent measurements in our entanglement-swapping
device. In the orthodox CHSH approach, Alice and Bob perform two
measurements on two copies of a given two-qubit state (four
measurements in total). However, to determine the Bell nonlocality
measure for an unknown state, they need to perform full QST
or to optimize their measurement bases, which requires performing
four measurements in each optimization step resulting in
many more measurements than in our experiment.

Our method solves the problem
of detecting and quantifying entanglement beyond a simple
Bell test in a typical entanglement-swapping 
method~\cite{Zukowski93PRL},  which can be applied to, e.g.,
quantum repeaters~\cite{Dur99PRA} and quantum
relays~\cite{Varnava16npjqi} in device-independent quantum
communications~\cite{Curty11PRA}, as well as to
entanglement-assisted quantum error
correction~\cite{Brun06Science} and entanglement
purification~\cite{Pan01Nat}.

We hope that our results could stimulate further research on
measuring such nonlinear properties of quantum systems as 
entanglement and nonlocality without performing full quantum-state
tomography.

\vspace*{3mm}
\noindent {\it Acknowledgements.---} KL and KB acknowledge the
financial support by the Czech Science Foundation under the
project No. 16-10042Y and the financial support of the Polish
National Science Centre under the grant No.
DEC-2013/11/D/ST2/02638. AČ acknowledges the financial support by
the Czech Science Foundation under the project No. P205/12/0382.
The authors also acknowledge the project No. LO1305 of the
Ministry of Education, Youth, and Sports of the Czech Republic
financing the infrastructure of their workplace. AM acknowledges
the support of a grant from the John Templeton Foundation.


%

\newpage

\end{document}


\title{Bell nonlocality and fully-entangled fraction measured
in an entanglement-swapping device without quantum state
tomography:\\ Supplemental Material}

\author{Karol Bartkiewicz} \email{bark@amu.edu.pl}
\affiliation{Faculty of Physics, Adam Mickiewicz University,
PL-61-614 Pozna\'n, Poland} \affiliation{RCPTM, Joint Laboratory
of Optics of Palack\'y University and Institute of Physics of
Czech Academy of Sciences, 17. listopadu 12, 771 46 Olomouc, Czech
Republic}

\author{Karel Lemr}
\email{k.lemr@upol.cz} \affiliation{RCPTM, Joint Laboratory of
Optics of Palack\'y University and Institute of Physics of Czech
Academy of Sciences, 17. listopadu 12, 771 46 Olomouc, Czech
Republic}

\author{Antonín Černoch} \email{acernoch@fzu.cz}
\affiliation{Institute of Physics of the Czech Academy of
Sciences, Joint Laboratory of Optics of PU and IP AS CR, 17.
listopadu 50A, 772 07 Olomouc, Czech Republic}

\author{Adam Miranowicz} \email{miran@amu.edu.pl}
\affiliation{Faculty of Physics, Adam Mickiewicz University,
PL-61-614 Pozna\'n, Poland}

\date{\today}

\begin{abstract}
Here we describe additional experimental details including the
maximum likelihood method, which ensures the positivity of the
reconstructed correlation matrix $R$, the directly measured
matrices (including the separable state $\ket{VV}$), and
other methods related to the inseparable  Werner states. We
also plotted the measured values of the relevant entanglement
witnesses.
\end{abstract}

\date{\today}
\maketitle

\subsection*{Experimentally measured matrices}

The experimentally obtained matrices
$\mathcal{R}^{(\mathrm{exp})}\equiv R^{(\mathrm{exp})}\pm\delta
R^{(\mathrm{exp})}$ for the assorted states read as
\begin{eqnarray}\nonumber
\mathcal{R}^{(\mathrm{exp})}_{\mathrm{sep}}&=&
\begin{bmatrix}
 \overline{.099} \pm .108 & \overline{.088} \pm .109 & \overline{.124} \pm .109 \\
 \cdots & \overline{.034} \pm .108 & \overline{.113} \pm .108 \\
 \cdots & \cdots & .980 \pm .147 \\
\end{bmatrix},\\\nonumber
\mathcal{R}^{(\mathrm{exp})}_{\mathrm{mix}}&=&
\begin{bmatrix}
 .017 \pm .031 & .006 \pm .031 & \overline{.007} \pm .031 \\
 \cdots & .013 \pm .033 & .016 \pm .033 \\
 \cdots & \cdots & .006 \pm .029 \\
\end{bmatrix},\\\nonumber
\mathcal{R}^{(\mathrm{exp})}_{\mathrm{ent}}&=&
\begin{bmatrix}
 .990 \pm .115 & .077 \pm .087 & .008 \pm .087 \\
 \cdots & .985 \pm .110 & \overline{.013} \pm .110 \\
 \cdots & \cdots & .959 \pm .079 \\
\end{bmatrix},
\end{eqnarray}
where $\overline{x}=-x$ and $\delta R^{(\mathrm{exp})}_{i,j}$ are
their experimental errors.

\subsection*{Maximum likelihood method}

To ensure the positivity of the reconstructed matrices, we
use the maximum likelihood method developed for quantum state
tomography (see, e.g.,~\cite{James01PRA}). We find the
physical matrix $R=[R_{i,j}]$, which is the closest to the
experimental but unphysical matrix
$R^{\mathrm{exp}}=[R^{(\mathrm{exp})}_{i,j}]$, by maximizing the
logarithmic likelihood function
\begin{equation}\label{eq:phys}
\mathcal{L}=-\sum_{1\le i\le j}^3 \left( \frac{R^{(\mathrm{exp})}_{i,j}-R_{i,j}}{\delta R^{(\mathrm{exp})}_{i,j}}\right)^2
\end{equation}
subject to $0\le r_j \le1$ for $j=1,2,3$ and
$[r_1,r_2,r_3]=\mathrm{eig}(R)$. This condition is equivalent to
requiring the probabilities of coincidence  detections to be
defined properly in any basis. Our maximum likelihood
estimates read as
$$R_{\mathrm{sep}}\approx
\begin{bmatrix}
 .008  &  .008  &  \overline{.086}  \\
 \cdots & .008  &  \overline{.091}  \\
 \cdots & \cdots & .982  \\
\end{bmatrix},\,
R_{\mathrm{mix}}\approx
\begin{bmatrix}
 .018  &  .004  &  \overline{.004}  \\
 \cdots & .015  &  .011  \\
 \cdots & \cdots & .010  \\
\end{bmatrix},$$
$$R_{\mathrm{ent}}\approx
\begin{bmatrix}
 .963  &  .038  &  .010  \\
 \cdots & .961  &  \overline{.012}  \\
 \cdots & \cdots & .959 \\
\end{bmatrix}.
$$
The corresponding spectra calculated for the exact maximum
likelihood estimates are $\mathrm{eig}(R_{\mathrm{mix}})=[0.019,
0.000, 0.024]$, $\mathrm{eig}(R_{\mathrm{ent}})= [0.919, 1.000,
0.965]$ and $R_{\mathrm{sep}}=[0.000, 0.998, 0.000]$.

The matrices $R$ are shifted on average by a fraction of
$0.19,\,0.02,\,0.07$ of $\delta R^{(\mathrm{exp})}$ from
$R^{(\mathrm{exp})}$ for the pure separable, maximally mixed, and
singlet states, respectively. Thus, we can assume that
$R^{(\mathrm{exp})}\approx R^{}$. The largest errors occur 
for the pure states. This is because the state is aligned with
only one of the eigenstates for the measurement apparatus. In
this case, we observe relatively low coincidence rates for 
the other eigenstates of the apparatus. Each matrix element
$R_{i,j}$ depends on four projections onto eigenstates of
$\sigma_{i}\otimes\sigma_{j}$ performed simultaneously by Bob.

\begin{table*}
\caption{\label{tab:data}The experimentally and theoretically
obtained values of the Bell nonlocality measure $M$
[Eq.~(4)], entropic witness $E$ [Eq.~(6)], and
FEF $F$ [Eq.~(5)].}
\begin{ruledtabular}
\begin{tabular}{lcccccc}
Density matrix  & $M^{(\mathrm{experiment})}$ & $E^{(\mathrm{experiment})}$ & $F^{(\mathrm{experiment})}$ & $M^{(\mathrm{theory})}$ & $E^{(\mathrm{theory})}$ & $F^{(\mathrm{theory})}$\\
 $\ket{\Psi^-}\bra{\Psi^-}$ & $+0.965 \pm 0.043$ & $+0.942 \pm 0.022$ & $+0.970 \pm 0.031$ & $+1$ & $+1$ & $+1$\\
 $\ket{VV}\bra{VV}$ & $-0.002 \pm 0.043$ &  $-0.001\pm 0.022$ & $-0.001\pm 0.08$& $0$ & $0$ & $0$\\
 $I/4$ & $-0.957 \pm 0.043$ & $-0.478 \pm  0.022$ & $-0.353\pm 0.141$ & $-1$ & $-\tfrac{1}{2}$ & $-\tfrac{1}{2}$
\end{tabular}
\end{ruledtabular}
\end{table*}

\begin{figure}
\vspace*{3mm}
\includegraphics[width=7.25cm]{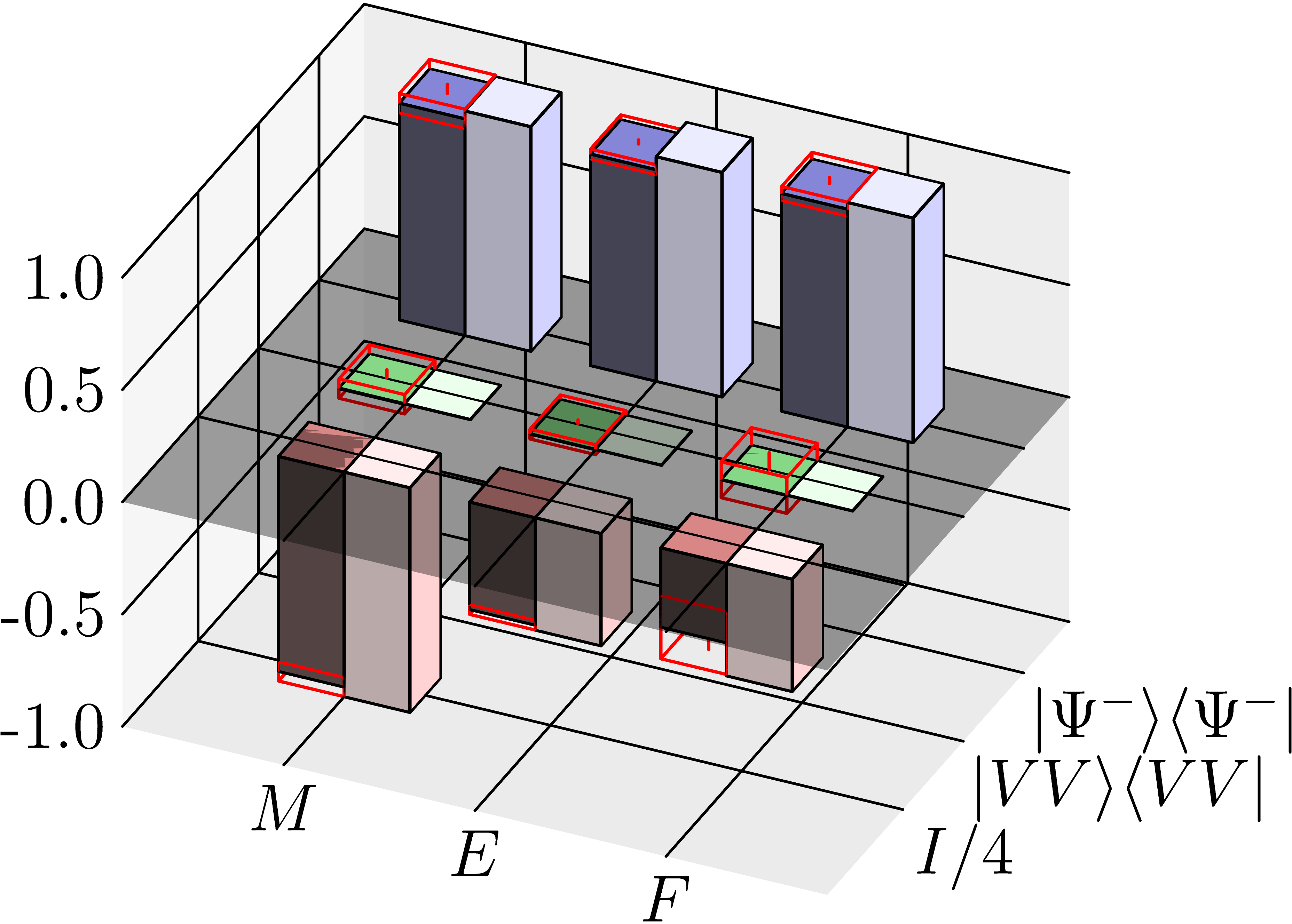}
\caption{\label{fig:bars} (Color online) Experimentally and
theoretically obtained values of the Bell nonlocality
measure $M$ [Eq.~(4)], the entropic witness $E$
[Eq.~(6)], and the FEF $F$ [Eq.~(5)]. The bright
(dark) bars correspond to theoretical (experimental) values. The
associated uncertainties are marked by red frames.}
\end{figure}

\subsection*{Measured entanglement witnesses}

Our maximum likelihood estimates were used to calculate the
values of the entanglement witnesses as summarized in
Fig.~\ref{fig:bars} and Tab.~\ref{tab:data}. The errors introduced
by the setup were estimated by comparing the results of the
measured etalon states ($\ket{VV}$ and $I/4$) with the theoretical
predictions.

\subsection*{The Werner states}

The spectrum of $R_{\mathcal{W}}$ matrix of the Werner states can
be expressed as $\mathrm{eig}(R_{\mathcal{W}}) \approx p^2
\mathrm{eig}(R_{\mathrm{ent}}) +(1-p)^2
\mathrm{eig}(R_{\mathrm{mix}}).$ This is approximation is valid if
$ \mathrm{eig}(R_{\mathrm{ent}})\gg
\mathrm{eig}(R_{\mathrm{mix}})\approx 0$ and the resulting
$\mathrm{Tr}\sqrt{R_{\mathcal{W}}}$ is linearly dependent on
the mixing parameter $p$ [for the ideal Werner states
$F=(3p-1)/2$].